\documentclass[smallabstract,smallcaptions]{dccpaper}

\usepackage{epsfig}
\usepackage{citesort}
\usepackage{color}
\usepackage{algorithm}
\usepackage{algorithmicx}
\usepackage{algpseudocode}
\usepackage{amsmath}
\usepackage{amssymb}
\usepackage{url}
\usepackage{comment}
\usepackage{multicol}

\newlength{\figurewidth}
\newlength{\smallfigurewidth}

\newtheorem{definition}{Definition}

\begin{document}

\title
{\large
\textbf{Fully Online Grammar Compression in Constant Space}\thanks{This work was supported by JSPS KAKENHI(24700140).}
}

\author{
Shirou Maruyama$^{\ast}$ and Yasuo Tabei$^{\dag}$ \\[0.5em]
{\small\begin{minipage}{\linewidth}\begin{center}
\begin{tabular}{ccc}
$^{\ast}$Preferred Infrastructure, Inc. & \hspace*{0.5in} & $^{\dag}$PRESTO, Japan Science\\
\url{maruyama@preferred.jp} && and Technology Agency \\
 && \url{tabei.y.aa@m.titech.ac.jp}
\end{tabular}
\end{center}\end{minipage}}
}

\maketitle
\thispagestyle{empty}

\begin{abstract}
We present novel variants of fully online LCA (FOLCA), a fully online grammar compression that builds a {\em straight line program} (SLP) and directly encodes it 
into a succinct representation in an online manner. 
FOLCA enables a direct encoding of an SLP into a succinct representation that is asymptotically equivalent to an information theoretic lower bound 
for representing an SLP (Maruyama et al., SPIRE'13). 
The compression of FOLCA takes linear time proportional to the length of an input text and its working space depends only on the size of the SLP, 
which enables us to apply FOLCA to large-scale repetitive texts. 
Recent repetitive texts, however, include some noise. 
For example, current sequencing technology has significant error rates, which embeds noise into genome sequences. 
For such noisy repetitive texts, FOLCA working in the SLP size consumes a large amount of memory. 
We present two variants of FOLCA working in constant space by leveraging the idea behind stream mining techniques. 
Experiments using 100 human genomes corresponding to about $300$GB from the 1000 human genomes project revealed the applicability of our method to large-scale, noisy repetitive texts. 
\end{abstract}

\section{Introduction}
Large-scale, highly repetitive text collections are becoming more and more ubiquitous. 
Examples are sequences of individual human genomes, source code in repositories and version controlled documents. 
In particular, the 1000 human genomes project stores more than 2,000 human genomes, which makes it possible for us to analyze human genetic variations~\cite{1000genomes}. 
Thus, we face a problem storing and processing a vast number of individual genome sequences of the same species, perhaps millions of genomes in the near future. 
There is therefore a strong demand for developing powerful methods to compress and process such text data on a large scale. 

Grammar compression is effective for compressing and processing repetitive texts.  
There are two types of problems in the field: (i) building a small context-free grammar (CFG) that generates a single text and (ii) representing the obtained CFG as compactly as possible. 
The method also reveals a processing ability for repetitive texts, e.g., pattern matching~\cite{Yamamoto2011}, pattern mining~\cite{Goto13} and edit distance computation~\cite{Hermelin09}. 
As far as we know, existing methods use two-step approaches: they first build a CFG from an input text and then encode it into a compact representation.
Although these methods achieve a high compression ratio, they require a large working space consumed by building the CFG and its encoding. 
Even worse, they are not applicable to streaming texts because of the indirectness of these steps. 

Maruyama et al.~\cite{Maruyama2013} overcame the efficiency problem by introducing a {\em fully online grammar compression} called {\em FOLCA} 
that builds a {\em straight line program} (SLP), a normal form of a CFG, and directly encodes it into a succinct representation in an online manner.
To achieve a small working space, they also presented a space-efficient implementation of {\em reverse dictionary}, 
a crucial data structure for checking whether or not a production rule in an SLP already exists in execution.
FOLCA consumes a small working space that depends only on the size of the SLP and not on the size of the input text because of the directness of building the SLP and its encoding in an online manner. 
In addition, the size of the succinct representation of an SLP of $n$ variables by FOLCA is $\lg{n}+2n+o(n)$ bits, which is asymptotically equivalent to 
the information theoretic lower bound of $\lg{n}+n+o(n)$ bits presented in \cite{Tabei13}. 
The compression time and approximation ratio are $O(N)$ and $O(\lg^2N)$, respectively, for the length $N$ of an input text, which 
are the same as those of the previous version of FOLCA~\cite{Maruyama2012}. 
Such nice properties of FOLCA enable the compression of a large-scale repetitive text in a small working space depending on the SLP size. 

Recent repetitive texts, however, include some noise. 
For example, current sequencing technology has significant error rates, which embeds noise into genome sequences. 
Actually, there is a $9\%$ difference on average among sequences stored in the 1000 human genomes project, 
although it is said that the genetic difference between individual human genomes is less than $0.01\%$. 
Since FOLCA works in a space depending on the SLP size, applying it to such noisy repetitive texts requires a large working space. 
One can solve the problem by dividing the input text into several blocks and applying compressors, e.g., LZ77 and RePair, into each block, but 
the compression ratio is ruined as long-range repetitions are not captured~\cite{KNtcs12}. 
Since the amount of large-scale repetitive text data including noise is ever increasing, developing a scalable grammar compression using a constant working 
space remains a challenge. 

\begin{table}[bt]
\begin{center}
{\footnotesize
  \caption{Comparison with existing algorithms.  Here, $N$ is the length of the input string, $\sigma$ is the alphabet size, $n$ is the number of generated rules, and $\alpha$ is a parameter between $0$ and $1$
  (the load factor of hash tables). 
  The expected time complexities are due to the use of a hash function.}
}
\label{tbl:comparison}
\footnotesize
  \begin{tabular}{c|c|l}
  Compression time & Working space (bits) & Ref. \\ \hline
  $O(N/\alpha)$ expected & $(3+\alpha)n \lg (n+\sigma)$ & \cite{Maruyama2012} \\
  $O(N/\alpha)$ expected & $(\frac{11}{4}+\alpha) n \lg (n+\sigma)$ & \cite{Takabatake2012} \\
  $O(N\lg n)$ & $2n \lg n(1+o(1))+ 2n\lg \rho$ ($\rho \leq 2\sqrt{n}$) & \cite{Tabei13} \\ 
  $O(\frac{N\lg n}{\alpha\lg\lg n})$ expected & $(1+\alpha) n \lg (n+\sigma) + n(3+\lg(\alpha n)) $ & \cite{Maruyama2013} \\ 
\hline
\hline
  $O(N/\alpha)$ expected & constant & this study
  \end{tabular}
\end{center}
\vspace{-0.5cm}
\end{table}

In this paper, we present novel variants of FOLCA that are fully online grammar compressions working in constant space for large-scale, noisy repetitive texts. 
Our variants output a succinct representation of an SLP to a secondary storage device and keep a hash table for a reverse dictionary in a main memory; this accounts for most of the memory consumed by the hash table for compression. 
To conserve working space, our variants keep only frequent production rules in the hash table by leveraging the idea behind stream mining techniques~\cite{Karp03,DemaineLM02,MankuM12}. 
Since these frequent production rules can represent as many as possible of the same digrams, they are expected to help achieve a high compression ratio. 
In addition, we also present a decompression algorithm working in constant space from our succinct representation of an SLP to efficiently recover a large text. 
Our results and those of existing algorithms are summarized in Table~\ref{tbl:comparison}.

Experiments were performed on compressing and decompressing 100 human genomes from the 1000 human genomes project. 
The results show the applicability of our method to large-scale, noisy repetitive texts. 

\section{Preliminaries}
\subsection{Basic notation}
Let $\Sigma$ be a finite alphabet for the symbols forming input texts throughout this paper. 
All elements in $\Sigma$ are totally ordered. 
$\Sigma^*$ denotes the set of all strings over $\Sigma$, and $\Sigma^i$ denotes the set of all strings of length $i$.
The length of $w \in \Sigma^*$ is denoted by $|w|$, and the cardinality of a set $C$ is similarly denoted by $|C|$. 
$\cal X$ is a recursively enumerable set of variables with $\Sigma \cap \cal X = \emptyset$. 
A sequence of symbols from $\Sigma \cup {\cal X}$ is also called a string, 
and an ordered pair of symbols from $\Sigma \cup {\cal X}$ is called a digram. 
Strings $x$ and $z$ are said to be the prefix and suffix of the string $w = xyz$, respectively, 
and $x, y, z$ are called substrings of $w$. 
The $i$-th symbol of $w$ is denoted by $w[i]$ ($1 \le i \le |w|$). 
For integers $i, j$ with $1 \leq i \leq j \leq |w|$, the substring of $w$ from $w[i]$ to $w[j]$ is denoted by $w[i,j]$. 
$\lg n$ stands for $\log_2 n$. 
Let $N$ be the length of an input text, which can be variable in an online setting. 

\subsection{Grammar compression}
A CFG is a quadruple $G=(\Sigma, V, D, X_s)$ where $V$ is a finite subset of $\cal X$, 
$D$ is a finite subset of $V \times (V \cup \Sigma)^*$ of production rules, 
and $X_s \in V$ represents the start symbol.
$D$ is also called a {\em phrase dictionary}. 
Variables in $V$ are called nonterminals. 
We assume a total order over $\Sigma \cup V$. 
The set of strings in $\Sigma^*$ derived from $X_s$ by $G$ is denoted by $L(G)$. 
A CFG $G$ is called {\em admissible} if for any $X \in {\cal X}$ there is exactly one production rule $X\to \gamma \in D$ and $|L(G)| = 1$.
An admissible $G$ deriving a text $S$ is called a grammar compression of $S$. 
The size of $G$ is the total of the lengths of strings on the right hand sides of all production rules; it is denoted by $|G|$.
The problem of grammar compression is formalized as follows:
\begin{definition}[Grammar Compression] \label{def:gc}
Given a string $w \in \Sigma^*$, compute a small, admissible $G$ that derives only $w$.
\end{definition}
In the following, we assume the case $|\gamma|=2$ for any production rule $X\to \gamma$. 
This assumption is reasonable because any grammar compression with $n$ variables can be transformed into such a restricted CFG with at most $2n$ variables.

The parse tree of $G$ is represented as a rooted ordered binary tree such that internal nodes are labeled by variables in $V$ and the {\em yields}, i.e., 
the sequence of labels of leaves that are equal to $S$. 
In a parse tree, any internal node $Z \in V$ corresponds to the production rule $Z\to XY$, and it has a left child labeled $X$ and a right child labeled $Y$. 
Let $height(X_i)$ be the height of the subtree having the root $X_i$ in the parse tree.
We assume an SLP for the CFG as follows.
\begin{definition}\label{SLP}(Karpinski-Rytter-Shinohara~\cite{SLP}) 
An SLP is a grammar compression over $\Sigma\cup V$
whose production rules are formed by 
$X_k\to X_iX_j$, where $X_i,X_j\in\Sigma\cup V$ and $1\leq i,j < k\leq |V| + |\Sigma|$.
\end{definition}
Note that although our definition of an SLP is different from the original definition~\cite{SLP} in 
that our production rules derive only digrams, they are equivalent. 
In this paper, we use our definition for notational convenience. 

\subsection{Reverse dictionary}
A reverse dictionary $D^{-1}: (\Sigma \cup {\cal X})^2 \to {\cal X}$ is a mapping from a given digram to a nonterminal symbol. 
$D^{-1}$ returns a nonterminal $Z$ associated with a digram $XY$ if $Z\to XY \in D$; otherwise, it creates a new nonterminal symbol $Z' \notin V$ and returns $Z'$. 
For example, if we have a phrase $D=\{X_1\to ab, X_2\to cd\}$, then $D^{-1}(a,b)$ returns $X_1$, while $D^{-1}(b,c)$ creates a new nonterminal $X_3$ and returns it. 
We can implement a reverse dictionary using a chaining hash table that has a load factor $\alpha$. 
The hash table has $\alpha n$ entries and each entry stores a list of a triple $(X_k,X_i,X_j)$ for a production rule $X_k\to X_iX_j$. 
For the rule $X_k\to X_iX_j$, the hash value is computed from $X_i$ and $X_j$. Then, the list corresponding to the hash value is scanned to search for $X_k$. 
Thus, the expected access time is $O(1/\alpha)$. 
The space is $\alpha n \lg{(n+\sigma)}$ bits for the hash table and $3n\lg{(n+\sigma)}$ bits for the lists. 
Therefore, the total size is $n(3+\alpha)\lg{(n+\sigma)}$ bits.

\section{FOLCA in Compressed Space}
FOLCA builds a post-order SLP (POSLP) as a special form of an SLP that can be transformed into a post-order partial parse tree (POPPT):   
a partial parse tree whose internal nodes are post-ordered. 
FOLCA directly encodes a POSLP into a succinct representation of a POPPT. 
In the original FOLCA, the succinct representation of a POPPT is kept in a main memory and used as the reverse dictionary in combination with a hash table. 
Instead, we modified FOLCA as it outputs the succinct representation of a POPPT into a secondary storage device, 
and we implemented a reverse dictionary using a chaining hash table kept in a main memory. 
Thus, the hash table is a crucial data structure for large-scale applications of FOLCA, because it accounts for most of the working space. 
In the next section, we present two space-efficient versions of FOLCA created by reducing the space of the hash table.

\begin{figure*}[t]
\begin{center}
\includegraphics[width=1.0\textwidth]{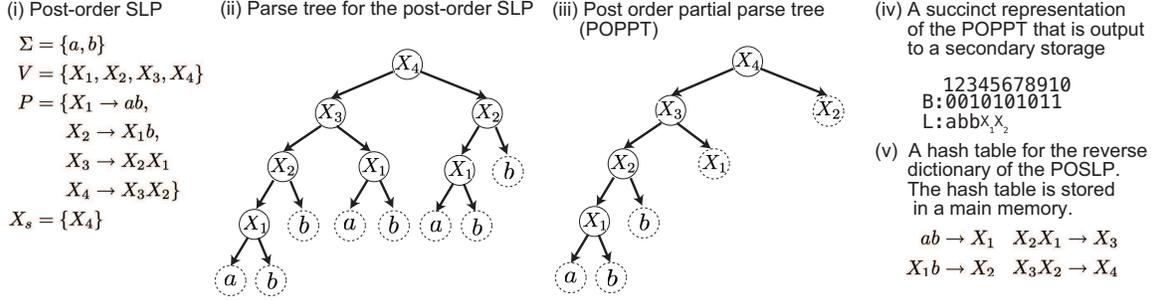}
\end{center}
\vspace{-0.6cm}
\caption{Example of a POSLP, parse tree of a POSLP, a post-order partial parse tree (POPPT), a succinct representation of POPPT and a hash table for a reverse dictionary. The succinct representation of the POPPT is output to a secondary storage device and the hash table is stored in a main memory.}
\label{fig:outline}
\end{figure*}

\subsection{Post-order SLP}
A partial parse tree defined by Rytter~\cite{Rytter03} is a binary tree formed by traversing a parse tree in a depth-first manner and pruning out all descendants under every node of nonterminal symbols appearing no less than twice. 
A POPPT and POSLP are defined as follows.
\begin{definition}[POSLP and POPPT~\cite{Maruyama2013}]
A post-order partial parse tree (POPPT) is a partial parse tree whose internal nodes have post-order variables. 
A post-order SLP (POSLP) is an SLP whose partial parse tree is a POPPT. 
\end{definition}

Note that the number of nodes in the POPPT is $2n + 1$ for a POSLP of $n$ variables, because the numbers of internal nodes  and leaves are $n$ and $n+1$, respectively.
Examples of a POSLP and POPPT are shown in Figure~\ref{fig:outline}-(i) and -(iii), respectively. 
In this example, all the descendants under every node having the second $X_1$ and $X_2$ in the parse tree (ii) are pruned out. 
The resulting POPPT (iii) has internal nodes consisting of post-order variables. 

A major advantage of the POSLP is that we can directly encode the corresponding POPPT into a succinct representation consisting of a bit string $B$ and a label sequence $L$. 
A bit string $B$ representing a POPPT is built by traversing a POPPT in a depth-first manner and placing '0' if a node is a leaf and '1' otherwise in post-order. The last bit '1' in $B$ represents a virtual node and is placed for fast tree operations without computing boundary conditions.
Thus, a POPPT of a POSLP of $n$ variables consists of $n$ internal nodes, $n+1$ leaves and a virtual node.
Label sequence $L$ keeps symbols in the leaves of a POPPT. 
The length of $L$ is $n+1$ and its space is $(n+1)\lceil \lg(n+\sigma) \rceil$ bits. 
Thus, the size of the POPPT of a POSLP of $n$ variables is $(n+1)\lceil \lg(n+\sigma)\rceil+2n+2$ bits. 

\subsection{FOLCA}
The basic idea of FOLCA is to (i) start from symbols of an input text, 
(ii) replace as many as possible of the same pairs of symbols in common substrings by the same nonterminal symbols, 
and (iii) iterate this process in a bottom-up manner until it generates a complete POSLP. 
The POSLP is built in an online fashion. 

FOLCA builds two types of subtrees in a POPPT from strings $XY$ and $XYZ$ of length two and three, respectively. 
The first type is a $2$-tree corresponding to a production rule in the form of $A\to XY$. 
The second type is a $2$-$2$-tree corresponding to production rules in the forms of $A\to YZ$ 
and $B\to XA$. 

To build a $2$-tree or $2$-$2$-tree from a substring of a limited length, FOLCA uses a {\em landmark}, which is defined as 
a local feature determined by strings of length four. 
Let $u$ be a string of length $m$. 
A function ${\mathcal L} : (\Sigma \cup V)^m \times [m] \to \{0,1\}$ classifies whether or not the $i$-th position of 
$u$ has a landmark, i.e., the $i$-th position of $u$ has a landmark if ${\mathcal L}(u,i)=1$. 
${\mathcal L}(u,i)$ is calculated from a substring $u[i-1]u[i][i+1]u[i+2]$ of length four. 
FOLCA builds a $2$-$2$-tree from a substring $u[i+1]u[i+2]u[i+3]$ of length three if the $i$-th position of $u$ does not have a landmark;  
otherwise, it builds a $2$-tree from a substring $u[i+2]u[i+3]$ of length two. 
Landmarks on a string are decided such that they are synchronized in long common subsequences  
to make the size of the POSLP generated by FOLCA as small as possible. 
See \cite{Maruyama2012} for the details of landmarks.

\begin{algorithm}[t]
{\footnotesize
\caption{FOLCA: Fully Online LCA.
$D$: phrase dictionary, $D^{-1}$: reverse dictionary, $q_k$: queue at level $k$.} 
\label{algo:FOLCA}
}
{\footnotesize
\begin{algorithmic}[1]
\Function{\sc FOLCA}{}
\State {$D:=\emptyset$; initialize queues $q_k$}
\While{Read a new character $c$ until it is not the end of the file}
\State{{\sc ProcessSymbol}($q_1$, $c$)}
\EndWhile
\EndFunction
\Function{{\sc ProcessSymbol}}{$q_k$, $X$}
\State $q_k.enqueue(X)$
\If {$q_k.size() = 4$}
\If {${\mathcal L}(q_k,2) = 0$} \Comment{Build a $2$-tree}
\State $Y :=$ {\sc Update}($q_k[3]$,$q_k[4]$)
\State {\sc ProcessSymbol}$(q_{k+1}, Y)$
\State $q_k.dequeue()$; $q_k.dequeue()$
\EndIf
\ElsIf {$q_k.size() = 5$} \Comment{Build a $2$-$2$-tree}
\State $Y :=$ {\sc Update}($q_k[4]$,$q_k[5]$); $Z :=$ {\sc Update}($q_k[3]$,$Y$)
\State {\sc ProcessSymbol}$(q_{k+1},Z)$
\State $q_k.dequeue()$; $q_k.dequeue()$; $q_k.dequeue()$
\EndIf
\EndFunction
\Function{{\sc Update}}{$X$,$Y$}
\State $Z := D^{-1}(X,Y)$
\If {$Z\to XY \notin D$}
\State $D:=D\cup \{Z\to XY\}$
\EndIf
\State return $Z$
\EndFunction
\end{algorithmic}
}
\end{algorithm}

The algorithm uses a set of queues, $q_k, k=1,...,m$, where $q_k$ corresponds to 
the $k$-th level of a parse tree of a POSLP and builds 2-trees and 2-2-trees at each level. 
Since FOLCA builds a balanced parse tree, the number $m$ of those queues is bounded by $\lg N$. 
In addition, landmarks are decided on strings of length four, and 
the lengths of these queues $q_k$, $k=1,...,m$, are also fixed to five. 
Algorithm~\ref{algo:FOLCA} consists of three functions, {\sc FOLCA}, {\sc ProcessSymbol} and {\sc Update}.
The main function is {\sc FOLCA}, which reads new characters from an input text and 
gives them to the function {\sc ProcessSymbol} one by one. 

The function {\sc ProcessSymbol} builds a POSLP in a bottom-up manner. 
There are two cases according to whether or not a queue $q_k$ has a landmark. 
For the first case of ${\mathcal L}(q_k,2)=0$, i.e., $q_k$ does not have a landmark, 
the $2$-tree corresponding to a production rule $Z\to q_k[3]q_k[4]$ in a POSLP
is built for the third and fourth elements $q_k[3]$ and $q_k[4]$ of the $k$-th queue $q_k$. 
For the other case,
the $2$-$2$-tree corresponding to production rules $Y\to q_k[4]q_k[5]$ and $Z\to q_k[3]Y$ is built for the third, fourth and fifth elements $q_k[3]$, $q_k[4]$ and $q_k[5]$ of the $k$-th queue $q_k$. 
In both cases, the function {\sc Update} returns a nonterminal symbol replacing a given digram. If the digram is a novel digram in execution, it returns a new nonterminal symbol, and otherwise, it returns the existing nonterminal symbol replacing the digram. 
The nonterminal symbol $Z$ is given to $q_{k+1}$ that is one level higher for a queue $q_{k}$, which enables the bottom-up construction of a POSLP in an online manner. 

The number of variables of a POSLP is $O(n_*\lg^2N)$ for the optimal grammar size $n_*$ and the length $N$ of an input text. 
The partial parse tree built by Algorithm~\ref{algo:FOLCA} is a POPPT. 
See \cite{Maruyama2013} for the proof. 

FOLCA builds nodes of the POPPT and directly encodes them into a succinct representation in an online manner. 
These nodes are output to a secondary storage device immediately after they have been built. 
Most of the working space is consumed by a hash table implementing a reverse dictionary~(Figure~\ref{fig:outline}-(v)). 
Therefore, the working space of FOLCA is at most $n(3+\alpha)\lg{(n+\sigma)}$ bits, while the compression time 
is $O(N/\alpha)$. 
In the next section, we introduce techniques for reducing the space of a hash table. 

\section{FOLCA in Constant Space}
We present novel variants of FOLCA working in constant space.
We reduce the size of the hash table implementing a reverse dictionary in FOLCA by leveraging the idea behind data mining techniques proposed in \cite{Karp03,DemaineLM02,MankuM12} for finding frequent items in data streams.
Our methods are {\em counter-based algorithms} that compute the frequencies of production rules of a dictionary in a streaming text and remove infrequent ones from the dictionary at a point in the algorithm. 
The frequent production rules in the resulting dictionary are expected to greatly contribute to string compression.
We present two algorithms called {\em frequency counting} and {\em lossy counting} that can be used instead of the function {\sc Update} in Algorithm~\ref{algo:FOLCA} to 
update a phrase dictionary and remove infrequent production rules.
We shall refer to FOLCAs using frequency counting and lossy counting as {\em FREQ\_FOLCA} and {\em LOSSY\_FOLCA}, respectively.



\subsection{FREQ\_FOLCA}
\begin{algorithm}[t]
{\footnotesize
\caption{Frequency counting. $k$: the maximum size of a phrase dictionary $D$, $\epsilon$: vacancy rate, $c$: frequency counter.
Note that {\sc FrequencyCounting} can be used instead of function {\sc Update} in Algorithm~\ref{algo:FOLCA}.} 
\label{algo:FREQUENT}
}
{\footnotesize
\begin{algorithmic}[1]
\Function{\sc FrequencyCounting}{$X$,$Y$}
\State $Z := D^{-1}(X,Y)$
\If {$Z\to XY \in D$}
\State {$c(Z) := c(Z) + 1$}
\Else
\If {$|D| \geq k$}
\While{$k(1-\epsilon/100) < |D|$}
\ForAll {$Z'\to X'Y' \in D$}
\State {$c(Z') := c(Z') - 1$}
\If {$c(Z') = 0$}
\State {$D := D \backslash \{Z' \rightarrow X'Y'\}$}
\EndIf
\EndFor
\EndWhile
\EndIf
\State{$D := D \cup \{Z\to XY\}$}
\State{$c(Z) := 1$}
\EndIf
\State\Return {$Z$}
\EndFunction
\end{algorithmic}
}
\end{algorithm}


The basic idea of frequency counting is to use a phrase dictionary of fixed-size $k$ 
and keep only frequent production rules in the phrase dictionary. 
Such frequently appearing production rules are expected to replace as many as possible of the same digrams by the same nonterminal symbols. 
Algorithm~\ref{algo:FREQUENT} shows an algorithm for frequency counting that can be used 
instead of function {\sc Update} in Algorithm~\ref{algo:FOLCA} to 
(i) generate a production rule, (ii) push it into a phrase dictionary and (iii) remove infrequent production rules. 
We use a frequency counter $c$ for each variable in a dictionary to compute its frequency. 
FREQ\_FOLCA builds a production rule $Z\to XY$ for a novel digram $XY$ and pushes it into a dictionary $D$ 
where the frequency counter $c(Z)$ for the left variable $Z$ is initialized to $1$. 
$c(Z)$ is incremented by one each time a digram $XY$ for $Z\to XY$ appears. 
If the dictionary size reaches $k$, 
FREQ\_FOLCA removes the bottom $\epsilon$ percent of infrequent production rules for a {\em vacancy rate} $\epsilon$. 

FREQ\_FOLCA generates one succinct POPPT each time {\sc FrequencyCounting} is called for an efficient decompression. 
Thus, each succinct POPPT represents a subtext of an input text. 
FREQ\_FOLCA recovers the original text for every succinct POPPT, which enables decompression using 
the same working space for compression. 

When a reverse dictionary is implemented using a chaining hash table that has a load factor $\alpha$, 
the working space of FREQ\_FOLCA is $k(\alpha+3)\lg(k+\sigma)$ bits, while 
the compression time is expected to be $O(N/\alpha)$.

\subsection{LOSSY\_FOLCA}
\begin{algorithm}
{\footnotesize
\caption{Lossy counting. $\ell$: parameter, 
$N$: length of an input string at a point in time, 
$X,Y \in (V \cup \Sigma)$. 
Note that {\sc LossyCounting} can be used instead of {\sc Update} in Algorithm~\ref{algo:FOLCA}.} 
\label{algo:LOSSYCOUNTING}
}
{\footnotesize
\begin{algorithmic}[1]
\State{Initialize $\Delta:=0$}
\Function{\sc LossyCounting}{X,Y}
\State $Z := D^{-1}(X,Y)$
\If{$Z\to XY \in D$}
\State {$c(Z) := c(Z) + 1$}
\Else
\State {$D := D \cup \{Z\to XY\}$}
\State {$c(Z) := \Delta + 1$}
\EndIf
\If {$\lfloor \frac{N}{\ell} \rfloor \neq \Delta$}
\State{$\Delta := N/\ell$}
\ForAll{$Z' \rightarrow X'Y' \in D$}
\If{$c(Z') < \Delta$}
\State{$D := D \backslash \{Z'\to X'Y'\}$}
\EndIf
\EndFor
\EndIf
\State\Return{$Z$}
\EndFunction
\end{algorithmic}
}
\end{algorithm}

The basic idea of lossy counting is to divide an input string into intervals of fixed length, 
and keep production rules in the next successive intervals according to the number of appearances of these production rules.
Thus, if a production rule appears no less than the total number of intervals, it is kept until the end of the execution.

Algorithm~\ref{algo:LOSSYCOUNTING} shows a lossy counting algorithm.
An input string is divided into $\ell$ intervals. 
Thus, the length of each interval is $N/\ell$. 
We use a counter $c$ for each variable for counting the number of appearances of  
each variable $Z$ . 
For a variable $Z$ appearing for the first time, $c(Z)$ is initialized to $N/\ell+1$, 
which means that a production rule $Z$ is kept in a dictionary for at least the next interval. 
If an existing variable $Z$ in a dictionary appears $r$ times, $c(Z)$ is incremented by $r$, resulting in $Z$ being kept in a dictionary for the next $r$ successive intervals. 
LOSSY\_FOLCA generates one succinct POPPT each time {\sc LossyCounting} is called for an efficient decompression just like FREQ\_FOLCA. 

The working space of LOSSY\_FOLCA is at most $2\ell(\alpha + 3)\lg(2\ell + \sigma)$ bits as a result of implementing a reverse dictionary as a chaining hash table, 
while the compression time is expected to be $O(N/\alpha)$. 

We also implemented another variant of FOLCA as a baseline by fixing $\Delta=\infty$ in Algorithm~\ref{algo:LOSSYCOUNTING}, which 
corresponds to dividing an input text into several blocks and applying FOLCA to each block. 
We shall call this variant of FOLCA {\em BLOCK\_FOLCA}. 
Comparing FREQ\_FOLCA and LOSSY\_FOLCA with BLOCK\_FOLCA reveals that BLOCK\_FOLCA using such a standard strategy for compressing large-scale 
texts cannot capture long-range repetitions, which is presented in Section~\ref{lab:experiments}.

\section{Decompression in Constant Space}
\begin{algorithm}
{\footnotesize
\caption{Decompression. $B$: bit string representing a POPPT,
$L$: label sequence of leaves in a POPPT, $D$: phrase dictionary, 
$S$: stack, 
$c$: counter for '0' and '1', 
$i$: counter for nonterminal symbols.}
\label{algo:DECOMPRESSION}
}
{\footnotesize
\begin{algorithmic}[1]
\State{Initialize $c:=0$; $D:=\emptyset$; $S:=\emptyset$; $i:=0$}
\Function{\sc Decompress}{$B$,$L$}
\While{Read $b$ from $B$}
\If{$b=0$} \Comment{$b$ is a leaf}
\State{$c:=c+1$}
\State{Read $A$ from $L$}
\State{Push $A$ to $S$}
\Else \Comment{$b$ is an internal node}
\State{$c:=c-1$}
\State{Pop $A$ and $B$ from $S$}
\State{$D \leftarrow \{X_i\to AB\}$}
\State{Push $X_i$ to $S$}
\State{$i:=i+1$}
\EndIf
\If{$c=0$}
\State{Pop $A$ from $S$}
\State{Recover subtext using $A$ and $D$}
\State{Remove infrequent production rules from $D$} \Comment{This step depends on the space reductions of hash table.}
\EndIf
\EndWhile
\EndFunction
\end{algorithmic}
}
\end{algorithm}

We recover the original text from multiple succinct POPPTs that FREQ\_FOLCA and LOSSY\_FOLCA output into a secondary storage device. 
Our succinct representation of POPPTs consists of a bit string $B$ and a label sequence $L$ presented in Figure~\ref{fig:outline}-(iv). 
We build a phrase dictionary $D$ by simulating a depth-first traversal which 
we can perform by gradually reading a bit from string $B$ and a label corresponding to a leaf label from $L$ from the beginning. 
Since a substring corresponding to one POPPT in $B$ has the same numbers of $'0'$ and $'1'$, respectively, 
we can detect each substring corresponding to a POPPT in $B$ by counting the number of $'0'$ and $'1'$. 
When we reach a position of $B$ corresponding to one POPPT, we recover the original string corresponding to a substring in $B$ 
from an obtained phrase dictionary $D$. 

The working spaces for decompression depend on the compression method. 
They are the same for FREQ\_FOLCA; likewise, they are the same for LOSSY\_FOLCA. 
The decompression time is $O(N/\alpha)$.

\section{Experiments} \label{lab:experiments}
\subsection{Setup}
We evaluated the performances of FREQ\_FOLCA and LOSSY\_FOLCA by comparing them with FOLCA, BLOCK\_FOLCA, and LZMA
 on one core of a quad-core Intel Xeon CPU E5-2680 (2.8GHz). 
We implemented FREQ\_FOLCA, LOSSY\_FOLCA, BLOCK\_FOLCA, and FOLCA in $C++$. 
LZMA is a general string compressor effective for repetitive texts, and 
we used its implementation downloadable from \url{http://sourceforge.jp/projects/sfnet_p7zip/}. 
We also used 100 human genomes corresponding to $306$GB downloadable from the 1000 human genomes project.
We tried $k=\{1000\mbox{MB}, 2000\mbox{MB}\}$ and $\epsilon=0.3$ for FREQ\_FOLCA and $\ell=\{5000\mbox{MB},10000\mbox{MB}\}$ for LOSSY\_FOLCA. 

\subsection{Results}
\begin{figure}[t]
\begin{center}
\begin{tabular}{cc}
compression & decompression \\
\includegraphics[width=0.45\textwidth]{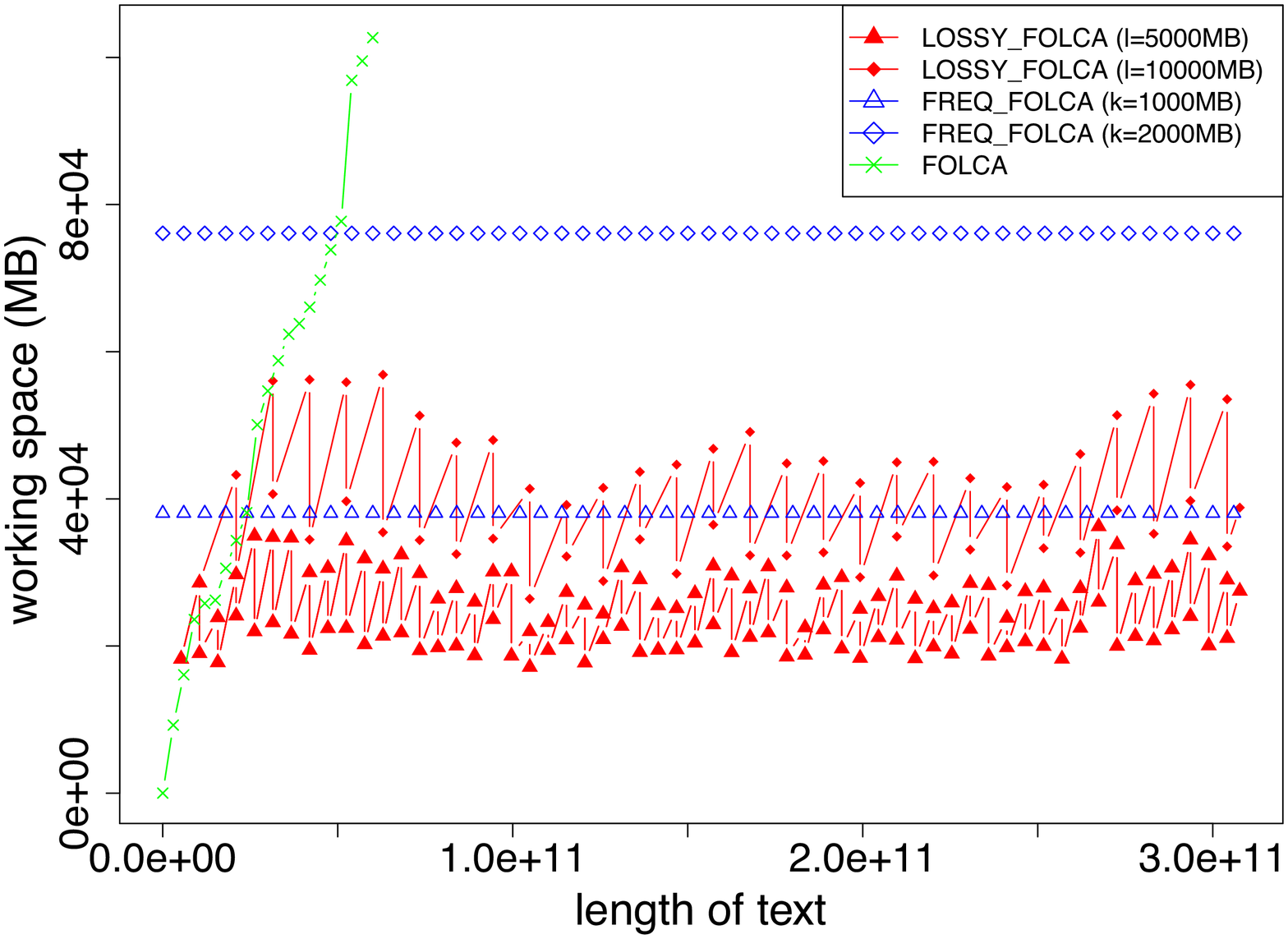} &
\includegraphics[width=0.46\textwidth]{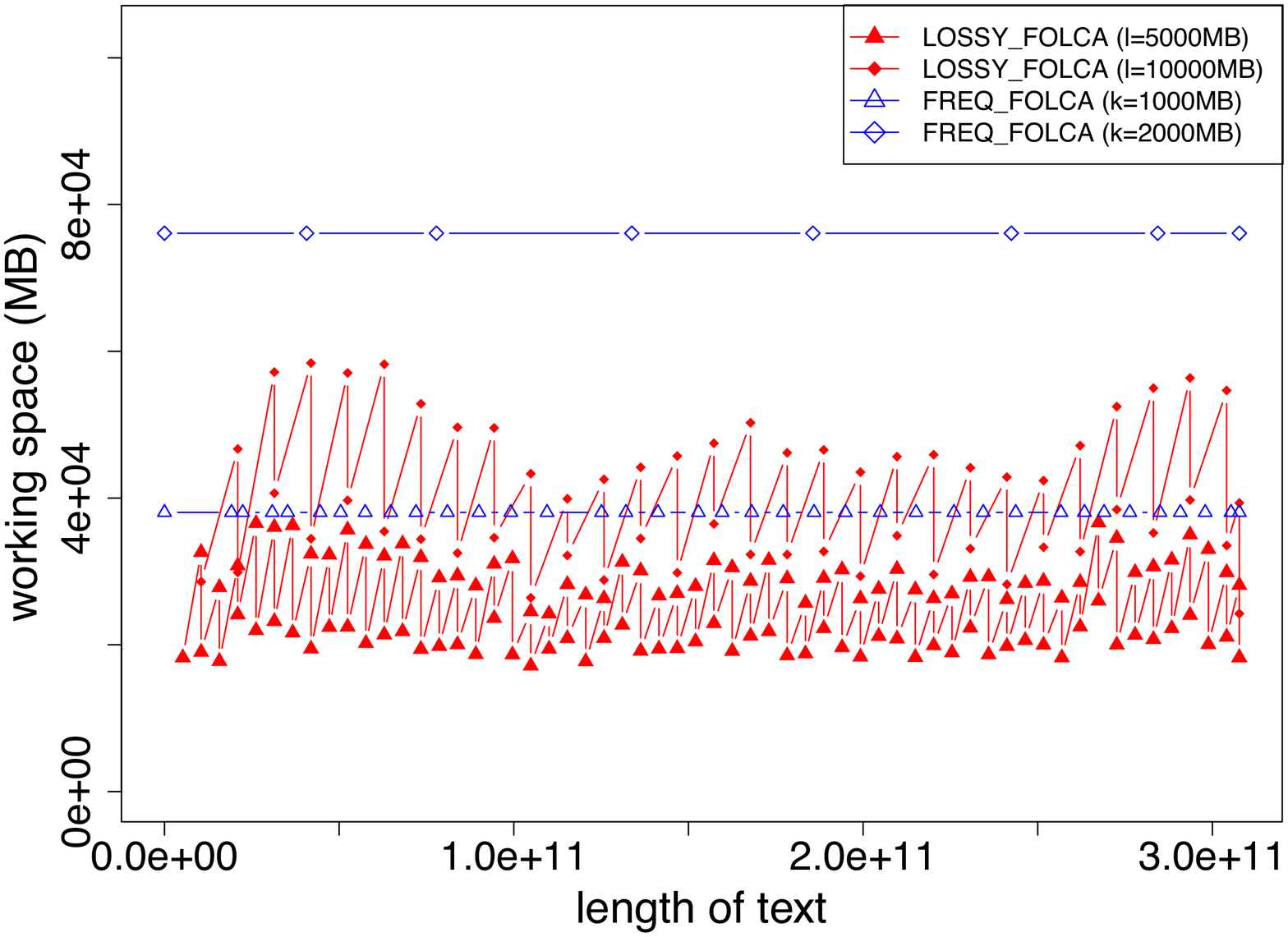} 
\end{tabular}
\end{center}
\vspace{-0.6cm}
\caption{Working space for the length of text in compression (left) and decompression (right).}
\label{fig:numspace}
\end{figure}

Figure~\ref{fig:numspace} shows the working space of each method for various lengths of a genome sequence in compression and decompression. 
FOLCA took more than $100$GB of working space for a genome sequence of $52$GB, which demonstrates that FOLCA, whose working space depends on 
the POSLP size, was not applicable to large-scale, noisy repetitive texts. 
LZMA did not finish within $5$ days, which shows
that LZMA is also not applicable to large-scale genome sequences. 
For both FREQ\_FOLCA and LOSSY\_FOLCA, the working spaces for compression and decompression were almost the same, despite parameter values being different.
The working space of FREQ\_FOLCA remained constant for compression and decompression: 
$36$GB and $76$GB for $k=1000$MB and $k=20000$MB, respectively. 
The working space of LOSSY\_FOLCA fluctuated, and 
the maximum values for $\ell=5000$MB and $\ell=10000$MB were $36$GB and $56$GB, which did not depend on the text length and SLP size.

Table~\ref{tab:result} shows the compression ratio, maximum working space for compression and decompression and compression/decompression time for 100 human genomes. 
A tradeoff between compression time and working space was observed for each method. 
The compression ratio for a larger working space was better for each method. 
The compression ratio of LOSSY\_FOLCA was better than that of BLOCK\_FOLCA for the same parameter value $\ell$, which showed that 
the strategy of LOSSY\_FOLCA for removing infrequent production rules was more effective than that of BLOCK\_FOLCA. 
While LOSSY\_FOLCA using a small working space achieved a higher compression ratio than FREQ\_FOLCA, 
one can use FREQ\_FOLCA when one wants to conserve working space. 
Compression using LOSSY\_FOLCA and FREQ\_FOLCA finished within about $24$hours. 
These results demonstrate the applicability of LOSSY\_FOLCA and FREQ\_FOLCA to large-scale, noisy repetitive texts. 

\begin{table}[th]
\begin{center}
{\scriptsize
\caption{Compression ratio (CR), compression time~(CT) in seconds~(s), decompression time (DT) in seconds and maximum working space~(WS) in megabytes~(MB) for each method on $100$ genome sequences.}
\label{tab:result}
}
\begin{tabular}{|l|r|r|r|r|}
\hline
Method & CR & CT~(s) & DT~(s) & WS~(MB) \\
\hline
FREQ\_FOLCA~(k=1000MB)   & $31.39$ & $86,098$ & $18,658$ & $38,048$ \\
FREQ\_FOLCA~(k=2000MB)   & $19.71$ & $93,823$ & $19,442$ & $76,096$ \\
LOSSY\_FOLCA~($\ell$=5000MB)  & $20.07$ & $87,548$ & $22,103$ & $36,246$ \\
LOSSY\_FOLCA~($\ell$=10000MB) & $17.45$ & $87,446$ & $20,378$ & $56,878$ \\
BLOCK\_FOLCA~($\ell$=5000MB)  & $31.85$ & $88,501$ & $-$ & $23,276$ \\
BLOCK\_FOLCA~($\ell$=10000MB) & $25.91$ & $92,007$ & $-$ & $34,665$ \\
\hline
\end{tabular}
\end{center}
\end{table}


\section{Conclusion}
We have presented fully-online grammar compressions working in constant space. 
Experimental results using 100 human genomes from the 1,000 human genome project demonstraite the applicability of our method to large-scale, noisy repetitive texts. 
Extensions are to develop various text processing methods applicable to large-scale repetitive texts. 
This would be beneficial to users for compressing and processing repetitive texts. 

\section{References}
\bibliographystyle{IEEEbib}
\bibliography{main.bib}

\begin{thebibliography}{10}

\bibitem{1000genomes}
1000 Genomes~Project Consortium,
\newblock ``A map of human genome variation from population-scale sequencing,''
\newblock {\em Nature}, vol. 467, pp. 1061--1073, 2010.

\bibitem{Yamamoto2011}
T.~Yamamoto, H.~Bannai, S.~Inenaga, and M.~Takeda,
\newblock ``Faster subsequence and don't-care pattern matching on compressed
  texts,''
\newblock in {\em Proceedings of the 22nd Annual Symposium on Combinatorial
  Pattern Matching}, 2011, vol. 6661, pp. 309--322.

\bibitem{Goto13}
K.~Goto, H.~Bannai, S.~Inenaga, and M.~Takeda,
\newblock ``Fast q-gram mining on {SLP} compressed strings,''
\newblock {\em Journal of Discrete Algorithms}, vol. 18, pp. 89--99, 2013.

\bibitem{Hermelin09}
D.~Hermelin, G.M. Landau, S.~Landau, and O.~Weimann,
\newblock ``A unified algorithm for accelerating edit-distance computation via
  text-compression,''
\newblock in {\em Proceedings of the 26th International Symposium on
  Theoretical Aspects of Computer Science}, 2009, pp. 529--540.

\bibitem{Maruyama2013}
S.~Maruyama, Y.~Tabei, H.~Sakamoto, and K.~Sadakane,
\newblock ``Fully-online grammar compression,''
\newblock in {\em Proceedings of the 20th String Processing and Information
  Retrieval Symposium}, 2013, pp. 218--229.

\bibitem{Tabei13}
Y.~Tabei, Y.~Takabatake, and H.~Sakamoto,
\newblock ``A succinct grammar compression,''
\newblock in {\em Proceedings of the 24th Annual Symposium on Combinatorial
  Pattern Matching}, 2013, pp. 218--229.

\bibitem{Maruyama2012}
S.~Maruyama, H.~Sakamoto, and M.~Takeda,
\newblock ``An online algorithm for lightweight grammar-based compression,''
\newblock {\em Algorithms}, vol. 5, pp. 213--235, 2012.

\bibitem{KNtcs12}
S.~Kreft and G.~Navarro,
\newblock ``On compressing and indexing repetitive sequences,''
\newblock {\em Theoretical Computer Science}, vol. 483, pp. 115--133, 2013.

\bibitem{Takabatake2012}
Y.~Takabatake, Y.~Tabei, and H.~Sakamoto,
\newblock ``Variable-length codes for space-efficient grammar-based
  compression,''
\newblock in {\em Proceedings of the 19th edition of the International
  Symposium on String Processing and Information Retrieval}, 2012, pp.
  398--410.

\bibitem{Karp03}
R.~Karp, S.~Shenker, and C.~Papadimitriou,
\newblock ``A simple algorithm for finding frequent elements in sets and
  bags,''
\newblock {\em ACM Transactions on Database Systems}, vol. 28, pp. 51--55,
  2003.

\bibitem{DemaineLM02}
D.~Demaine, A.~L{\'o}pez-Ortiz, and I.~Munro,
\newblock ``Frequency estimation of internet packet streams with limited
  space,''
\newblock in {\em Proceedings of the 10th European Symposium on Algorithms},
  2002, pp. 348--360.

\bibitem{MankuM12}
G.~Manku and R.~Motwani,
\newblock ``Approximate frequency counts over data stream,''
\newblock in {\em Proceedings of the 28th International Conference on Very
  Large Data Bases}, 2002, vol.~5, pp. 346--357.

\bibitem{SLP}
M.~Karpinski, W.~Rytter, and A.~Shinohara,
\newblock ``An efficient pattern-matching algorithm for strings with short
  descriptions,''
\newblock {\em Nordic Journal of Computing}, vol. 4, pp. 172--186, 1997.

\bibitem{Rytter03}
W.~Rytter,
\newblock ``Application of {Lempel-Ziv} factorization to the approximation of
  grammar-based compression,''
\newblock {\em Theoretical Computer Science}, vol. 302, no. 1-3, pp. 211--222,
  2003.

\end{thebibliography}

\end{document}